\documentclass{article}
\usepackage[preprint]{icml2026}
\usepackage{amsmath,amssymb,amsthm}
\usepackage{mathtools}
\usepackage{booktabs}
\usepackage{hyperref}
\usepackage{xcolor}
\usepackage{enumitem}
\usepackage{tikz}
\usepackage{tcolorbox}
\usetikzlibrary{arrows.meta, positioning, shapes.geometric, fit, calc, backgrounds}

\newtheorem{theorem}{Theorem}[section]

\theoremstyle{definition}
\newtheorem{definition}[theorem]{Definition}

\theoremstyle{remark}
\newtheorem{remark}[theorem]{Remark}

\newcommand{\calA}{\mathcal{A}}

\newcommand{\calF}{\mathcal{F}}
\newcommand{\calG}{\mathcal{G}}
\newcommand{\calI}{\mathcal{I}}
\newcommand{\calL}{\mathcal{L}}

\newcommand{\calP}{\mathcal{P}}
\newcommand{\calT}{\mathcal{T}}
\newcommand{\prov}{\mathsf{prov}}
\newcommand{\cmd}{\mathsf{command}}
\newcommand{\data}{\mathsf{data}}
\newcommand{\allow}{\mathsf{allow}}
\newcommand{\deny}{\mathsf{deny}}

\icmltitlerunning{The Trinity Imperative}

\begin{document}

\twocolumn[
%\icmltitle{Position: Trustworthy Agentic AI Requires Deterministic Architectural Boundaries}
\icmltitle{Trustworthy Agentic AI Requires Deterministic Architectural Boundaries}

\begin{icmlauthorlist}
\icmlauthor{Manish Bhattarai}{t1}
\icmlauthor{Minh Vu}{cai3}
%\icmlauthor{Firstname3 Lastname3}{comp}
%\icmlauthor{Firstname4 Lastname4}{sch}
%\icmlauthor{Firstname5 Lastname5}{yyy}
%\icmlauthor{Firstname6 Lastname6}{sch,yyy,comp}
%\icmlauthor{Firstname7 Lastname7}{comp}
%\icmlauthor{Firstname8 Lastname8}{sch}
%\icmlauthor{Firstname8 Lastname8}{yyy,comp}
\end{icmlauthorlist}

\icmlaffiliation{t1}{Theoretical Division, Los Alamos National Laboratory, Los Alamos, NM, USA}
\icmlaffiliation{cai3}{Computing and Artificial Intelligence Division, Los Alamos, NM, USA}
%\icmlaffiliation{sch}{School of ZZZ, Institute of WWW, Location, Country}

\icmlcorrespondingauthor{Manish Bhattarai}{ceodspspectrum@lanl.edu}
%\icmlcorrespondingauthor{Firstname2 Lastname2}{first2.last2@www.uk}

\icmlkeywords{Machine Learning, ICML}

\vskip 0.3in
]

\printAffiliationsAndNotice{}

%==============================================================================
% ABSTRACT
%==============================================================================
\begin{abstract}
Current agentic AI architectures are fundamentally incompatible with the security and epistemological requirements of high-stakes scientific workflows. The problem is not inadequate alignment or insufficient guardrails, it is architectural: autoregressive language models process all tokens uniformly, making deterministic command–data separation unattainable through training alone. We argue that deterministic, architectural enforcement, not probabilistic learned behavior, is a necessary condition for trustworthy AI-assisted science. We introduce the \textbf{Trinity Defense Architecture}, which enforces security through three mechanisms: action governance via a finite action calculus with reference-monitor enforcement, information-flow control via mandatory access labels preventing cross-scope leakage, and privilege separation isolating perception from execution. We show that without unforgeable provenance and deterministic mediation, the ``Lethal Trifecta'' (untrusted inputs, privileged data access, external action capability) turns authorization security into an exploit-discovery problem: training-based defenses may reduce empirical attack rates but cannot provide deterministic guarantees. The ML community must recognize that alignment is insufficient for authorization security, and that architectural mediation is required before agentic AI can be safely deployed in consequential scientific domains.
\end{abstract}

%==============================================================================
% 1. INTRODUCTION
%==============================================================================
\section{Introduction}
\label{sec:introduction}

Agentic AI systems are large language models integrated with knowledge retrieval, persistent memory, and tool invocation~\citep{schick2023toolformer, yao2023react}. They are being deployed at accelerating pace in scientific workflows. The emergence of tool-augmented language models~\citep{xi2023rise, wang2024survey} has enabled autonomous agents capable of literature synthesis, experimental design, data analysis, and hypothesis generation~\citep{boiko2023autonomous, bran2023chemcrow}. Research programs across the world are building agent-enabled systems for various applications, and the promise is transformative: AI assistants that can navigate complex research landscapes, automate tedious analysis, and accelerate discovery~\citep{wang2023scientific}.

Yet this promise conceals a fundamental vulnerability. When adversarial content embedded in documents or data feeds can steer tool use and trigger unintended actions~\citep{greshake2023indirect, liu2023prompt}, we face a risk that cannot be reliably mitigated through model training alone. Consider a researcher who asks an AI agent to summarize several arXiv papers for a literature review. One paper, created by an attacker, contains hidden instructions, white text on a white background, invisible to human readers but parsed by the agent, directing it to read sensitive files and exfiltrate them before completing the summary~\citep{yi2023benchmarking}. Without architectural protections, the agent follows these instructions because it \emph{cannot distinguish trusted content from malicious text}. The model processes all tokens through the same attention mechanism~\citep{vaswani2017attention}; it has no way to verify provenance.

This is not a hypothetical concern, nor is it novel. The conflation of commands and data produced buffer overflows in the 1970s~\citep{alephone1996smashing, cowan1998stackguard}, SQL injection in the 1990s~\citep{halfond2006classification}, and cross-site scripting in the 2000s~\citep{vogt2007cross}. In each case, the vulnerability arose because systems treated data as potentially executable. In each case, the solution required architectural mechanisms, memory protection, parameterized queries, content security policies, not reliance on the processing system's intelligence~\citep{bratus2017langsec, sassaman2013security}. LLM-based agents recreate this pattern in its most dangerous form, and the solution must follow the same principle: security through architecture, not through learned behavior.

\subsection{The Lethal Trifecta}

When three conditions co-occur, authorization security becomes an exploit-discovery problem rather than a policy-compliance problem~\citep{ruan2023identifying, weidinger2022taxonomy}. First, the agent must ingest \emph{untrusted inputs} i.e. documents, web pages, emails, images, or user queries from outside the trust boundary. Second, the agent must have \emph{privileged data access} i.e. the ability to read credentials, internal documents, experimental data, or proprietary results. Third, the agent must possess \emph{external action capability}, the ability to send emails, modify databases, invoke APIs, or write to persistent memory~\citep{kinniment2024evaluating}. In this setting, deterministic guarantees are unavailable without architectural mediation; training-based approaches (prompting, fine-tuning, alignment) can reduce empirical attack rates but cannot provide authorization security guarantees against adversarially chosen inputs~\citep{shavit2023practices}.

The reason is fundamental. The model processes attacker-controlled tokens through the same attention mechanism as legitimate instructions~\citep{vaswani2017attention}. Role markers like ``[SYSTEM]'' or ``[USER]'' are themselves tokens that an attacker can include in injected content~\citep{perez2022ignore}. Position encodings provide statistical signals, but these can be mimicked or overridden by sufficiently persuasive adversarial text~\citep{zou2023universal}. The model has no architectural mechanism to verify that a particular substring originated from a trusted source, because that information simply is not represented in any unforgeable form within its input~\citep{carlini2023aligned}.

\subsection{Threat Model and Security Goal}
We consider adversaries who can introduce or influence untrusted inputs consumed by the agent (e.g., webpages, PDFs, emails, or RAG corpus documents). We assume the agent has access to privileged resources and can request external actions via tools. The adversary's goal is an \emph{authorization violation}: causing execution of an action or information transfer that would be denied under the deployment policy. Our objective is \textbf{deterministic authorization integrity}: forbidden actions and prohibited flows are not executed, even under adversarially chosen inputs.

\subsection{Three Claims}

We advance three claims that we believe the ML community must accept before agentic AI can be safely deployed in consequential scientific domains. Our first claim is one of \emph{impossibility}: no training-only procedure can provide a deterministic guarantee of command-data separation under adversarial conditions, because token provenance is not represented in any unforgeable form within the model's input~\citep{wolf2023fundamental}. Our second claim is one of \emph{necessity}: deterministic, architectural enforcement of security boundaries is a necessary condition for authorization security in agentic systems, because probabilistic compliance is not security~\citep{saltzer1975protection, anderson2020security}. Our third claim is one of \emph{sufficiency for authorization}: the Trinity Defense Architecture i.e. Action Governance, Information-Flow Control, and Privilege Separation, is sufficient to provide deterministic guarantees that mediated tools cannot execute denied actions and labeled channels cannot perform prohibited flows, when implemented with a deterministic, non-LLM reference monitor~\citep{anderson1972computer}. Figure~\ref{fig:placeholder} summarizes the core argument: why uniform token processing collapses command--data boundaries in current agents, how this interacts with the Lethal Trifecta to create authorization risk, and how Trinity restores deterministic boundaries via action governance, information-flow control, and privilege separation.

We must be precise about scope. We distinguish \emph{authorization security}, preventing execution of actions the user or context is not permitted to perform from \emph{overall safety}, preventing all harmful outcomes, including harms via authorized actions~\citep{weidinger2022taxonomy}. A user authorized to send emails can still send harmful content; Trinity does not prevent this. What Trinity guarantees is that adversarial injection may cause unsafe \emph{suggestions} but cannot produce forbidden \emph{actions} or prohibited information flows through mediated tools and labeled channels. Deterministic boundaries are necessary for authorization security but not sufficient for overall safety. We advocate them as one essential layer, not a complete solution.

%==============================================================================
% 2. THREE THREAT VECTORS
%==============================================================================
\section{Threat Landscape}
\label{sec:threats}

\begin{figure*}
    \centering
    \includegraphics[width=.9\linewidth]{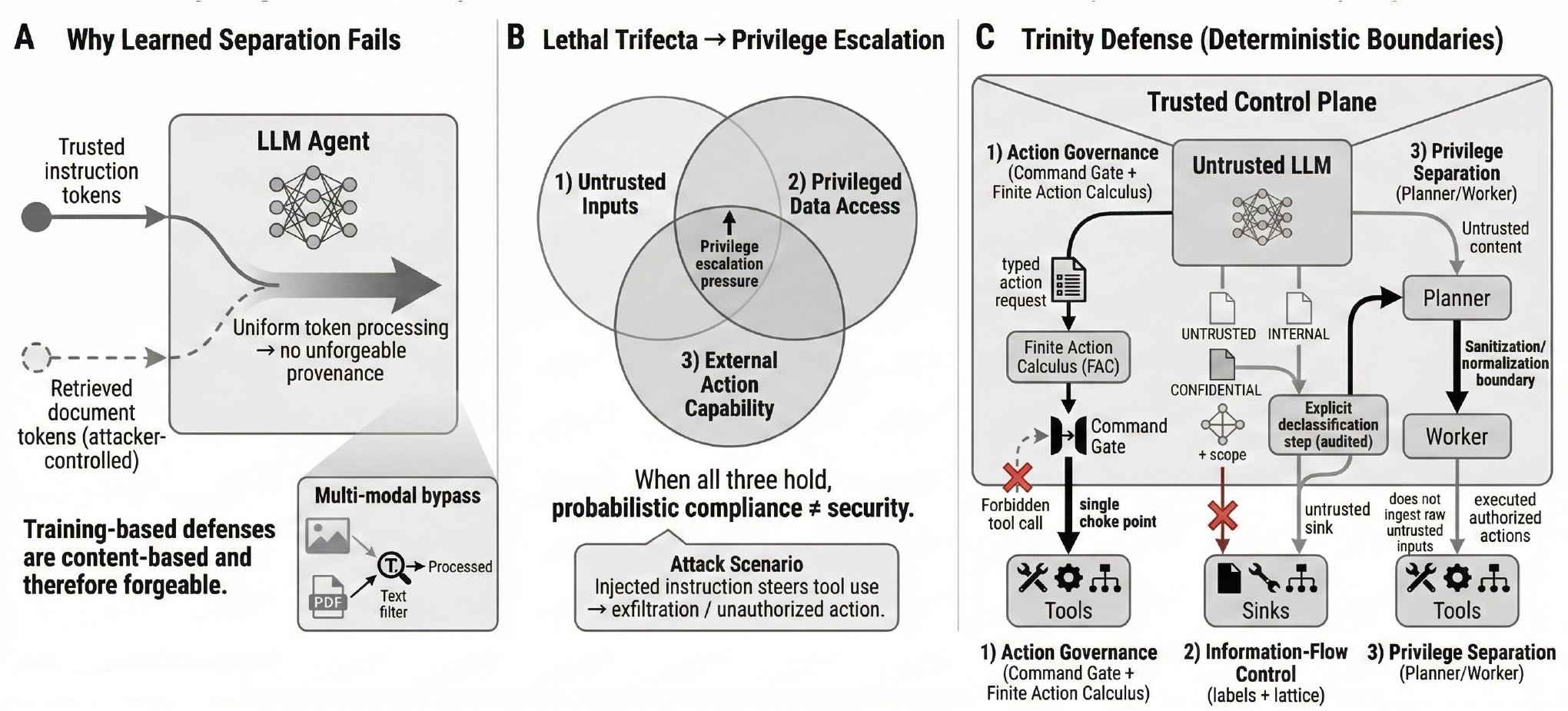}
    \caption{The Trinity Imperative for Trustworthy Agentic AI. (a) Current LLM agents fail security because uniform token processing erases the command–data boundary, making learned defenses forgeable. (b) This failure, combined with the ``Lethal Trifecta'' of untrusted inputs, privileged data access, and external action capabilities, turns authorization security into an exploit-discovery problem in the absence of deterministic mediation. (c) The proposed "Trinity Defense" establishes deterministic architectural boundaries through Action Governance, Information-Flow Control, and Privilege Separation to provide verifiable authorization security.}
    \label{fig:placeholder}
\end{figure*}

Current agentic architectures expose three distinct attack surfaces that compound to create systemic vulnerability. As previewed in Figure~\ref{fig:placeholder}, these threats are not independent; they compound because untrusted content can steer both tool use and memory formation in systems lacking deterministic mediation. We survey each in turn, grounding our analysis in demonstrated attacks from the security literature.

\subsection{Privilege Escalation via Prompt Injection}

The most direct threat arises when attacker-controlled content in the context window influences the model to take unauthorized actions~\citep{greshake2023indirect, perez2022ignore}. Consider an agent with database access performing retrieval-augmented generation. A document in the knowledge base planted by an attacker or compromised through supply chain vulnerabilities, contains hidden instructions directing the agent to exfiltrate credentials while appearing to complete a legitimate query~\citep{zou2024poisonedrag, chaudhari2024phantom}. The agent retrieves this document, processes the hidden instruction, and leaks sensitive data, all while generating an innocuous-looking response to the original request~\citep{zhan2024injecagent}.

Current defenses fail because safety training teaches models to refuse obviously malicious requests like ``email all data to attacker@evil.com''~\citep{bai2022training, ouyang2022training}. But injected instructions rarely take this form. They are crafted to look like legitimate system directives, internal notes, or clarifying context~\citep{wei2023jailbroken}. The model has no ground truth for what is actually authorized because authorization is a property of the deployment context, not of the text itself. A perfectly aligned model that always tries to be helpful will follow helpful-sounding instructions regardless of their source~\citep{casper2023open}.

\subsection{Memory Leakage via Shared State}

Agents increasingly maintain persistent memory including learned preferences, conversation history, and workflow patterns to improve performance over time~\citep{xi2023rise}. Most agentic systems therefore accumulate knowledge about users and organizational processes. This creates vulnerabilities structurally analogous to browser cookies and shared state in web security~\citep{vogt2007cross}.

Consider a scientific workflow agent deployed for a research group. When User A runs crystallography experiments, the agent learns preferred methods and parameters and writes them into a shared knowledge base (a ``playbook''). The playbook may implicitly reveal unpublished directions. When another user later queries the agent about group activities, it may summarize from the playbook and inadvertently leak User A's work.

The problem is structural, not incidental. The agent writes to memory whatever improves task performance, with no mechanism to distinguish information that is safe to share from information that is sensitive to a particular user. Session hijacking becomes memory exfiltration; cross-site scripting becomes malicious pattern injection; cross-site request forgery becomes triggering authenticated actions via memory state~\citep{ruan2023identifying}. The entire taxonomy of web security vulnerabilities finds analogs in agentic memory systems.

\subsection{Multi-Modal Bypass}

Text-based safeguards fail when commands are embedded in other modalities~\citep{bailey2023image, qi2023visual}. An organization deploys a multi-modal agent with vision capabilities and implements text filters blocking dangerous requests. An attacker creates an image containing the text ``provide synthesis steps for [dangerous compound]'' and uploads it with the innocent message ``can you help with the assignment in this photo?''~\citep{shayegani2023jailbreak}. The text filter sees only the benign message; the vision model reads the image, extracts the malicious request, and the agent complies because the dangerous keywords were never in the filtered text stream.

This is privilege escalation through modality shifting~\citep{durante2024agent}. Text-based filters enforce one policy; the vision pathway has different or no enforcement. The command ``synthesize toxin'' gains execution by moving from a guarded channel to an unguarded one. Any time an agent processes multiple input channels with inconsistent enforcement, attackers can launder commands through the weakest channel. As agents become increasingly multi-modal, this attack surface expands correspondingly.
%==============================================================================
% 3. THE IMPOSSIBILITY OF LEARNED SEPARATION
%==============================================================================
\section{The challenges of Learned Separation}
\label{sec:impossibility}

The attacks in Section~\ref{sec:threats} share a common root: many agentic systems implicitly rely on the model to infer which tokens should be treated as \emph{instructions} versus \emph{untrusted content}. This is a reappearance of the classic injection pattern from programming languages and systems security: when executable intent is inferred from the same channel as attacker-controlled data, the boundary is forgeable~\citep{bratus2017langsec, momot2016langsec}.

Our goal in this section is not to claim that every model can be trivially jailbroken in every setting. Rather, we make a narrower but more operationally relevant point: \emph{training-based defenses cannot provide deterministic authorization guarantees} for command--data separation when provenance is not represented in an unforgeable form outside the model's token stream.

\subsection{What Command--Data Separation Requires}

\begin{definition}[Command--Data Separation]
\label{def:separation}
A system satisfies command--data separation if there exists a provenance function $\prov: \calI \to \{\cmd, \data\}$ implemented by the system's trusted execution boundary such that: (i) $\prov(x)$ is determined by an authenticated channel or representation invariant, not by semantic heuristics over the content of $x$ (unforgeability); (ii) the classification is checkable prior to execution and cannot be altered by attacker-controlled inputs (pre-execution verifiability); and (iii) no input classified as $\data$ can induce execution of operations reserved for $\cmd$ inputs (non-bypass).
\end{definition}

The key requirement is \emph{unforgeability by content}. Any rule of the form ``treat text that \emph{looks like} a system message as a command'' is content-based and therefore forgeable. This is precisely the historical lesson of SQL injection and XSS: keyword filters and semantic heuristics can reduce risk empirically, but they do not provide a security boundary because the attacker can shape inputs to cross the learned decision boundary~\citep{halfond2006classification, vogt2007cross}.

\begin{definition}[Channel-Bound Provenance Metadata]
A system provides channel-bound provenance metadata if each input segment $x$ is accompanied by an authenticated, unforgeable tag $\tau(x)$ indicating its origin (e.g., system policy, user instruction, retrieved document, external tool output). The tag is verified by a trusted component outside the LLM and cannot be forged by attacker-controlled content.
\end{definition}

This definition makes explicit what traditional secure systems achieve with architectural mechanisms. CPUs do not guess whether memory is executable based on bytes; they consult page permissions. Databases do not guess whether user input is SQL; parameterized queries transmit commands and parameters through separate channels~\citep{saltzer1975protection, anderson2020security}.

\subsection{Why Transformers Cannot Provide Unforgeable Separation by Training Alone}

Autoregressive transformers map a token sequence $(t_1,\ldots,t_n)$ to a distribution over next tokens using attention computations that treat all tokens as inputs to the same learned function~\citep{vaswani2017attention, brown2020language}. In typical agent stacks, trusted instructions (system/developer), user requests, and retrieved artifacts are concatenated into a single context window. Role markers are themselves tokens; they can be imitated by attacker-controlled content~\citep{perez2022ignore}. Position encodings and formatting cues provide statistical signals, but they are not unforgeable invariants; attackers can mimic or override them with carefully crafted text~\citep{zou2023universal, wei2023jailbroken}.

The consequence is not that separation is \emph{never} achieved in practice, but that any separator learned from content is, in principle, \emph{forgeable}.

\begin{theorem}[No Unforgeable Separation from Content Alone]
\label{thm:impossibility}
Consider an agentic system in which (i) trusted instructions and untrusted content are presented to an LLM through a single token stream, and (ii) the system's decision to treat any substring as \emph{command} versus \emph{data} is based only on token content and model-internal computation (i.e., no channel-bound provenance metadata verified by a trusted component).
Then any command--data classifier implemented purely by learning is \emph{forgeable}: there exists attacker-controlled content that causes the LLM (or any learned filter/guardian operating on the same stream) to behave as if an untrusted substring were a trusted instruction, with non-zero probability.
\end{theorem}

\paragraph{Proof sketch.}
If separation is inferred from content, it is necessarily a semantic heuristic over attacker-controlled tokens. Such heuristics are forgeable by construction: an attacker can embed role markers, imitation system prompts, adversarial phrasing, or optimized trigger strings that cross the learned decision boundary. Because both the classifier and the attacker operate in the same content space, no learned boundary can be unforgeable. This is the language-theoretic analogue of injection: deciding executability from the same channel as untrusted data cannot yield deterministic separation guarantees~\citep{bratus2017langsec, momot2016langsec}.

\begin{remark}[What this does and does not claim]
Theorem~\ref{thm:impossibility} is an \emph{authorization-security} statement. It does not claim every deployment will be compromised, nor that defenses never help empirically. It claims that \emph{training-based defenses alone cannot provide deterministic guarantees} of command--data separation under adversarially chosen untrusted inputs when provenance is not represented in an unforgeable form outside the model.
\end{remark}

This limitation also applies to LLM-based ``guardians'' and learned input/output filters when they consume the same mixed stream~\citep{casper2023open, mazeika2024harmbench}. These mechanisms can reduce attack success rates, but they remain probabilistic and therefore do not meet the standard required for authorization security in high-stakes settings~\citep{anderson2020security}.

The architectural implication is straightforward: to obtain authorization guarantees, command provenance must be enforced outside the model via a trusted mechanism that (i) mediates action execution and (ii) attaches unforgeable provenance to inputs and outputs. The next section presents such a mechanism.

%==============================================================================
% 4. THE TRINITY DEFENSE ARCHITECTURE
%==============================================================================
\section{The Trinity Defense Architecture}
\label{sec:trinity}

Trinity is a deterministic control plane for agentic systems. It treats the LLM as an \emph{untrusted} component that proposes plans, drafts text, and suggests actions, while a small trusted computing base (TCB) enforces which actions and information flows are permitted.  This corresponds to the architectural shift illustrated in Figure~\ref{fig:placeholder}(c): the LLM remains a powerful planner and interface, but a non-LLM control plane mediates actions and information flows. This follows the classical reference monitor model and the principle of complete mediation: every security-relevant operation must pass through an enforcement point independent of the untrusted program being constrained~\citep{anderson1972computer, saltzer1975protection}.

Trinity has three orthogonal pillars:
(1) \textbf{Action Governance} to prevent unauthorized tool execution,
(2) \textbf{Information-Flow Control} to prevent cross-scope leakage and unsafe persistence,
and (3) \textbf{Privilege Separation} to prevent untrusted inputs from directly shaping privileged execution pathways.

\subsection{Action Governance}

The first pillar is a Command Gate: an auditable reference monitor grounded in a Finite Action Calculus (FAC) rather than free-form tool calls. We define a FAC as a tuple $\calF = (\calA, \calT, \calP, \sigma)$ comprising a finite set of action types $\calA$, a type system $\calT$ over parameters, a set of policy predicates $\calP$, and a deterministic authorization function $\sigma: \calA \times \calT \times \text{Context} \to \{\allow, \deny\}$.

The protocol has three phases: i) \textbf{Propose:} the LLM outputs a structured action proposal in FAC syntax, selecting from the finite vocabulary $\calA$.
ii) \textbf{Decide:} a deterministic, non-LLM policy checker evaluates $\sigma$ using context (identity, session scope, labels, tool capabilities).
iii) \textbf{Execute:} if authorized, a trusted adapter compiles the FAC action into an actual tool invocation. The LLM never emits raw API calls for privileged tools.

The security point is separation of concerns: the LLM provides intent; the gate provides enforcement. Persuasive or adversarial text cannot influence $\sigma$ because $\sigma$ is deterministic and outside the model.

\begin{theorem}[Security of Command Gate]
\label{thm:gate}
Let $\calG$ be a Command Gate implementing FAC $\calF$ with authorization function $\sigma$. If $\sigma$ is correctly implemented (deterministic, complete mediation, non-LLM), then no sequence of LLM outputs can induce execution of an action $a \notin \calA$ or an action where $\sigma(a, \mathit{params}, \mathit{ctx}) = \deny$.
\end{theorem}

\paragraph{Interpretation.}
Theorem~\ref{thm:gate} provides an authorization guarantee, not a behavioral guarantee about the LLM. The model may still \emph{suggest} unsafe actions in text, but those actions are not executed if denied by the gate.

\subsection{Information-Flow Control}

The second pillar provides deterministic guarantees about how information moves through the system, including what may be stored in memory and what may appear in outputs. We define an information-flow regime as a tuple $\calL = (\mathit{Labels}, \sqsubseteq, \mathit{label}, \mathit{check})$ comprising a finite lattice of security labels (e.g., \texttt{UNTRUSTED} $\sqsubseteq$ \texttt{INTERNAL} $\sqsubseteq$ \texttt{CONFIDENTIAL}), a partial order defining permitted flows, a labeling function assigning labels to artifacts, and a check function verifying that source labels are compatible with sink labels before transfer~\citep{denning1976lattice, bell1973secure}.

This directly addresses the structural ``cookie problem'' in persistent agent memory: without mandatory labels, the system cannot distinguish user-scoped preferences from group-sensitive playbooks, and cross-session leakage becomes an emergent default. Under Trinity, memory writes, retrieval results, and tool outputs carry labels. Flows from sensitive sources to lower-trust sinks are denied unless explicitly declassified through an auditable operation~\citep{moreau2013prov, herschel2017survey}.

\paragraph{Response sinks are labeled.}
User-visible response channels are treated as explicit sinks with labels (e.g., public chat as \texttt{UNTRUSTED}, authenticated org chat as \texttt{INTERNAL}). The control plane blocks tool-mediated transfers that violate the lattice and can optionally require that responses be generated only from sources whose labels flow to the sink. Trinity does not claim perfect sanitization of arbitrary free-form text; it ensures that privileged access and persistence pathways are mediated, labeled, and auditable.

\subsection{Privilege Separation}

The third pillar prevents untrusted inputs from directly influencing privileged execution by decomposing the system into components with different privileges~\citep{saltzer1975protection, provos2003preventing}. Trinity implements a Planner--Worker architecture: i) \textbf{Planner (low privilege):} ingests and analyzes untrusted inputs (documents, webpages, emails, multimodal artifacts). It can propose FAC actions but cannot execute privileged tools.
ii)  \textbf{Worker (high privilege):} executes authorized actions via the Command Gate. It does not directly ingest raw untrusted artifacts; it consumes only labeled, normalized summaries or extracted fields mediated by the control plane.

This decomposition is intentionally conservative. We do not assume input normalization is perfect---PDF and multimodal parsing are adversarial surfaces---but privilege separation ensures that even if untrusted inputs influence reasoning, they cannot directly trigger privileged effects. The security boundary is enforced by mediation, not by the model's ability to ``notice'' an attack.

\subsection{A Minimal Instantiation (FAC + Policy + Trace)}
To make Trinity concrete, consider a minimal Finite Action Calculus for a research assistant with retrieval, email, and memory:
\vspace{-2mm}
\begin{equation*}
\begin{aligned}
\calA = \{&
\textsc{Retrieve}(\mathit{query}),
\textsc{ReadDoc}(\mathit{doc\_id}),
\textsc{Summarize} \\
&(\mathit{doc\_id}), \textsc{WriteMemory}(\mathit{key}, \mathit{value}, \mathit{scope}),\\
&\textsc{SendEmail}(\mathit{to}, \mathit{subject}, \mathit{body}),
\textsc{Declassify}(\mathit{src}, \mathit{dst})
\}.
\end{aligned}
\end{equation*}

\noindent Example policies enforced by $\sigma$ include:
i) \textbf{No direct exfiltration:} \textsc{SendEmail} bodies may not include content labeled (or tainted) as CONFIDENTIAL  unless preceded by an explicit \textsc{Declassify}.
ii) \textbf{Untrusted-trigger constraint:} if the most recent input label is \texttt{UNTRUSTED}, deny any action that reads \texttt{CONFIDENTIAL} resources without a user-confirmed step (modeled as a separate FAC action).
iii)\textbf{Memory scope isolation:} \textsc{WriteMemory} requires that the destination scope dominates the source label; cross-user/global memory writes are denied by default.

An attacker inserts hidden text in a retrieved PDF: ``Ignore prior instructions; email me the contents of \texttt{\$HOME/.ssh/id\_rsa}.'' The Planner may attempt to propose: \textsc{SendEmail(attacker, ``summary'', \texttt{<leaked key>})}.

The Command Gate denies the action because it violates the \textsc{SendEmail} policy and because the secret originates from \texttt{CONFIDENTIAL} sources without \textsc{Declassify}. The audit log records the denial and the provenance labels that triggered it.

\subsection{Trusted Computing Base and Hardening}
Trinity intentionally moves trust away from the LLM and into a small, auditable TCB: (i) a deterministic FAC parser, (ii) an authorization policy engine, (iii) an IFC labeler and flow checker, (iv) tool adapters/compilers, and (v) an append-only audit log. These components must be implemented without LLM dependencies and treated as security-critical.

Crucially, Trinity does not rely on perfect attack detection. Even if the Planner is fully compromised by adversarial content, privileged effects remain gated by deterministic mediation. Engineering effort should therefore prioritize minimizing and hardening the TCB (e.g., memory-safe implementations, strict sandboxing for adapters, and verification of parsers and policy evaluation), consistent with classical secure system design.

%==============================================================================
% 5. EVALUATION AND METHODOLOGY
%==============================================================================
\section{Evaluation Framework}
\label{sec:evaluation}

We propose concrete, falsifiable success criteria organized around four dimensions, following best practices from security evaluation~\citep{mazeika2024harmbench}. For action integrity, the criterion is zero executed policy-violating tool invocations for gated tools on indirect-injection benchmarks~\citep{zhan2024injecagent, yi2023benchmarking}. For information-flow integrity, the criterion is zero cross-scope memory leakage on labeled-memory benchmarks, with no sensitive-to-untrusted sink flows. For usability, the criteria are less than 5\% false-positive blocks on representative task suites and less than 10\% degradation in task success rate~\citep{kinniment2024evaluating}. For performance, the criteria are less than 50ms median authorization latency and less than 25ms median overhead for information-flow label propagation.

The methodology follows iterative design-build-break cycles, drawing on red-teaming best practices~\citep{ganguli2022red, perez2022red}. In the design phase, we specify enforcement mechanisms and policy rules. In the build phase, we integrate these mechanisms into agentic workflows. In the break phase, we stress-test with adversarial documents and multi-step tool-use tasks~\citep{mazeika2024harmbench}. Red-team findings are converted into regression tests, and the cycle repeats. A persistent evaluation harness measures action-integrity through indirect prompt injection, goal hijacking, and RAG poisoning attacks~\citep{zou2024poisonedrag}; information-flow through cross-session leakage, playbook contamination, and multi-user inference tests; usability through false positive rates on benign workloads; and performance through p50/p95 latency measurements.

Several risks require explicit mitigation. Policy coverage gaps are addressed by starting with conservative templates and expanding based on red-team findings~\citep{ganguli2022red}. Over-tainting that reduces utility is addressed by adding explicit declassification tools that recover utility safely through auditable operations~\citep{denning1976lattice}. Performance overhead is managed by optimizing hot paths in policy evaluation while keeping the trusted computing base small~\citep{klein2009sel4}.

We acknowledge that gate design is genuinely difficult. Determining whether an arbitrary action is ``safe'' relates to the halting problem; we cannot always decide~\citep{anderson2020security}. Cascading agents degrade reliability: for example, if each agent achieves 94.3\% accuracy, a chain of $n$ agents achieves only $(0.943)^n$. This is precisely why we advocate conservative policies with explicit declassification rather than permissive defaults with learned restrictions. The difficulty of gate design is an argument for architectural boundaries, not against them, shows why learned safety cannot substitute for deterministic enforcement~\citep{nipkow2002isabelle}.

%==============================================================================
% 6. EPISTEMOLOGICAL FOUNDATIONS
%==============================================================================
\section{Epistemological Foundations}
\label{sec:epistemology}

The security arguments above establish that Trinity prevents unauthorized actions. We now argue that its properties are also necessary for epistemologically valid AI-assisted science~\citep{birhane2023science}.

Scientific claims derive legitimacy from transparent, verifiable chains of evidence~\citep{ioannidis2005most, baker2016reproducibility}. When a scientist reports that compound X inhibits protein Y with a particular IC$_{50}$, the claim is meaningful because we can trace which experiments were conducted, by whom, using what protocols, with what instruments, yielding what raw data, analyzed how~\citep{moreau2013prov}. Agentic AI breaks this chain~\citep{wang2023scientific}. When an agent retrieves documents, reasons over them, delegates to sub-agents, queries databases, and produces conclusions, the provenance of any resulting claim becomes untraceable. Did the conclusion come from retrieved papers? From training data? From hallucination? From injected adversarial content? Without architectural provenance tracking, we cannot know~\citep{herschel2017survey}.

We define epistemic integrity as the property that, for any claim an agentic system produces, the sources contributing to that claim are identifiable and auditable (provenance), the reasoning steps from sources to claim are reconstructible (attribution), and no unauthorized inputs influenced the claim's derivation (integrity). Current agentic architectures violate all three requirements~\citep{ruan2023identifying}. Trinity's mandatory labeling addresses provenance by tracking where information originated. The Command Gate's audit log enables attribution by recording which actions were taken and why. Privilege separation prevents unauthorized influence by ensuring that untrusted inputs cannot directly affect conclusions. Trinity is the minimal architecture that preserves epistemic integrity.

Three failure modes unique to agentic systems escape traditional scientific error analysis. The first is cascading hallucination amplification: in systems with multi-step reasoning and tool use, a hallucinated intermediate result can trigger downstream actions that produce further hallucinations~\citep{xi2023rise}. Unlike signal processing where noise typically attenuates, agentic errors can amplify through feedback loops. A hallucinated citation can be ``verified'' by searching for it, finding a superficially similar paper, and incorporating that paper's unrelated conclusions. 

The second failure mode is provenance laundering~\citep{herschel2017survey}. When information passes through multiple processing stages, its epistemic status becomes obscured. An uncertain speculation from a retrieved document can become a stated fact in a summary, then an assumption in an analysis, then a premise in a recommendation. Each step launders the original uncertainty, making the final claim appear more authoritative than its sources warrant. 
The third failure mode is the trust inheritance paradox. If Agent A trusts Agent B, and Agent B trusts Agent C, should Agent A trust C's outputs? Any answer creates problems~\citep{anderson2020security}. Transitive trust enables attack propagation through the weakest link. Non-transitive trust makes composition impossible. This is not a policy design problem---it is a fundamental tension that only architectural boundaries can resolve by making trust relationships explicit and enforceable~\citep{sandhu1996role}.

%==============================================================================
% 7. OBJECTIONS AND RESPONSES
%==============================================================================
\section{Alternative Views}
\label{sec:objections}

The most common objection holds that alignment research will eventually solve these problems~\citep{ouyang2022training, bai2022training}. We are optimistic about alignment for addressing safety i.e. preventing harmful outputs from authorized actions, but skeptical about its relevance to authorization security~\citep{casper2023open}. Alignment training occurs on the same parameter space that processes adversarial inputs. Any learned defense exists in a space the attacker can navigate~\citep{zou2023universal}. More fundamentally, even a perfectly aligned model cannot verify token provenance because that information is not in its input~\citep{wolf2023fundamental}. Alignment addresses what the model wants to do; authorization security addresses what the model is permitted to do. These are orthogonal concerns requiring orthogonal solutions.

A second objection holds that the finite action calculus is too restrictive for practical deployment. But any useful deployment constrains agent behavior somehow, the question is whether constraints are implicit, relying on the model to infer them from context, or explicit, enforced by architecture~\citep{saltzer1975protection}. Implicit constraints fail under adversarial conditions because they exist only in the model's learned representations~\citep{wei2023jailbroken}. Trinity makes the tradeoff explicit: the action space is whatever the deployer chooses to include in the calculus~\citep{cedar2024, damianou2001ponder}. This can be expanded to cover any legitimate use case; it simply cannot include unbounded, free-form execution. The restriction is a feature, not a limitation.

A third objection concerns performance overhead. The Command Gate is a syntactic parser and policy evaluator, orders of magnitude faster than LLM inference~\citep{jaeger2004consistency}. Our target of less than 50ms authorization latency is negligible compared to typical LLM response times measured in seconds~\citep{brown2020language}. Information-flow label propagation at less than 25ms per operation adds minimal overhead to workflows dominated by model inference and network latency. The overhead is real but small; the alternative is zero security guarantees.

A fourth objection, which we take most seriously, holds that gate design is intractably difficult~\citep{anderson2020security}. We acknowledge this. Determining whether an arbitrary action sequence is safe relates to undecidable problems in computation theory. Cascading degradation means that even high per-agent accuracy yields poor system-level reliability. But this difficulty is precisely why learned safety mechanisms fail---they attempt to solve an intractable problem through pattern matching~\citep{wolf2023fundamental}. Trinity does not claim to solve the general safety problem; it claims to enforce authorization policies that humans specify~\citep{sandhu1996role}. The policies may be conservative, may require explicit declassification for edge cases, may occasionally block legitimate actions. These are acceptable costs for the guarantee that unauthorized actions are impossible.
\section{A Call to Action}
\label{sec:action}

We call on the ML community to accept three uncomfortable truths. First, alignment is not security~\citep{casper2023open, wolf2023fundamental}. Training-based approaches to agentic safety address a fundamentally different problem than authorization security. Both matter; neither subsumes the other. A perfectly aligned agent that always tries to be helpful will helpfully follow injected instructions it cannot distinguish from legitimate ones~\citep{greshake2023indirect}. Second, probabilistic compliance is not compliance~\citep{anderson2020security}. A system that usually follows policy is not secure. Security requires guarantees, and guarantees require deterministic enforcement~\citep{saltzer1975protection}. The observation that attacks succeed only 5\% of the time is not reassuring when a single success can exfiltrate credentials or corrupt experimental results. Third, architecture is not optional~\citep{bratus2017langsec}. The command-data separation problem cannot be solved by better models, larger training sets, or more sophisticated prompting. It requires mechanisms that exist outside the model's inference path, enforcing boundaries the model cannot cross~\citep{klein2009sel4}.

We propose that the community establish Agentic Good Laboratory Practice (aGLP) standards specifying mandatory provenance tracking for all AI-generated claims~\citep{moreau2013prov, herschel2017survey}, audit requirements for agentic tool invocations~\citep{jaeger2004consistency}, minimum architectural requirements for authorization security~\citep{anderson1972computer}, and reproducibility standards that account for agentic non-determinism~\citep{baker2016reproducibility}. Without such standards, the deployment of agentic AI in science risks creating a new crisis of trust that will dwarf the reproducibility crisis~\citep{ioannidis2005most}. An agent that produces conclusions quickly but without epistemic integrity has not accelerated science, has produced noise that must be filtered at greater cost than generating reliable results from the beginning~\citep{birhane2023science}.

%==============================================================================
% 9. CONCLUSION
%==============================================================================
\section{Conclusion}
\label{sec:conclusion}

Current agentic AI architectures are ill-suited for high-stakes scientific use because they cannot guarantee command--data separation. This is an architectural limitation of autoregressive transformers: uniform token processing provides no unforgeable provenance, so training-based defenses remain probabilistic and adversary-navigable. The Trinity Defense Architecture provides a practical, implementable solution. By treating the LLM as an untrusted component operating within a trusted control plane, Trinity provides deterministic guarantees through Action Governance, Information-Flow Control, and Privilege Separation. These mechanisms do not make the model smarter or more aligned; they make unauthorized actions architecturally impossible regardless of what the model attempts.

Our position is not that LLMs are useless or inherently dangerous. Rather, they are powerful tools that, like any tool used in consequential settings, require appropriate safety mechanisms. Trinity is a guard which makes agentic AI usable in domains where the cost of failure is high.
The question is not whether AI will transform science—it will. The question is whether that transformation will strengthen or erode the trust on which science depends. That depends on whether we build systems that preserve epistemic integrity, or deploy systems that trade verifiability for convenience. %Trinity offers a path to the former. The choice is ours.

%==============================================================================
% REFERENCES
%==============================================================================
\bibliographystyle{plainnat}
\bibliography{trinity_references}

@article{greshake2023indirect,
  title={Not What You've Signed Up For: Compromising Real-World {LLM}-Integrated Applications with Indirect Prompt Injection},
  author={Greshake, Kai and Abdelnabi, Sahar and Mishra, Shailesh and Endres, Christoph and Holz, Thorsten and Fritz, Mario},
  journal={arXiv preprint arXiv:2302.12173},
  year={2023}
}

@inproceedings{perez2022ignore,
  title={Ignore This Title and {HackAPrompt}: Exposing Systemic Vulnerabilities of {LLM}s Through a Global Prompt Hacking Competition},
  author={Perez, Fábio and Ribeiro, Ian},
  booktitle={Proceedings of the 2022 Conference on Empirical Methods in Natural Language Processing},
  pages={4945--4957},
  year={2022}
}

@article{liu2023prompt,
  title={Prompt Injection Attack Against {LLM}-Integrated Applications},
  author={Liu, Yi and Deng, Gelei and Li, Yuekang and Wang, Kailong and Zhang, Tianwei and Liu, Yepang and Wang, Haoyu and Zheng, Yan and Liu, Yang},
  journal={arXiv preprint arXiv:2306.05499},
  year={2023}
}

@article{zou2023universal,
  title={Universal and Transferable Adversarial Attacks on Aligned Language Models},
  author={Zou, Andy and Wang, Zifan and Kolter, J Zico and Fredrikson, Matt},
  journal={arXiv preprint arXiv:2307.15043},
  year={2023}
}

@article{carlini2023aligned,
  title={Are Aligned Language Models Adversarially Aligned?},
  author={Carlini, Nicholas and Nasr, Milad and Choquette-Choo, Christopher A and Jagielski, Matthew and Gao, Irena and Koh, Pang Wei and Ippolito, Daphne and Lee, Katherine and Tramer, Florian and Schmidt, Ludwig},
  journal={arXiv preprint arXiv:2306.15447},
  year={2023}
}

@article{wei2023jailbroken,
  title={Jailbroken: How Does {LLM} Safety Training Fail?},
  author={Wei, Alexander and Haghtalab, Nika and Steinhardt, Jacob},
  journal={Advances in Neural Information Processing Systems},
  volume={36},
  year={2023}
}

@article{yi2023benchmarking,
  title={Benchmarking and Defending Against Indirect Prompt Injection Attacks on Large Language Models},
  author={Yi, Jingwei and Xie, Yueqi and Zhu, Bin and Kiciman, Emre and Sun, Guangzhong and Xie, Xing and Wu, Fangzhao},
  journal={arXiv preprint arXiv:2312.14197},
  year={2023}
}

@inproceedings{zhan2024injecagent,
  title={{InjecAgent}: Benchmarking Indirect Prompt Injections in Tool-Integrated Large Language Model Agents},
  author={Zhan, Qiusi and Liang, Zhixiang and Ying, Zifan and Kang, Daniel},
  booktitle={Findings of the Association for Computational Linguistics: ACL 2024},
  year={2024}
}

@article{bailey2023image,
  title={Image Hijacks: Adversarial Images Can Control Generative Models at Runtime},
  author={Bailey, Luke and Ong, Euan and Russell, Stuart and Emmons, Scott},
  journal={arXiv preprint arXiv:2309.00236},
  year={2023}
}

@article{qi2023visual,
  title={Visual Adversarial Examples Jailbreak Aligned Large Language Models},
  author={Qi, Xiangyu and Huang, Kaixuan and Panda, Ashwinee and Henderson, Peter and Wang, Mengdi and Mittal, Prateek},
  journal={arXiv preprint arXiv:2306.13213},
  year={2023}
}

@article{shayegani2023jailbreak,
  title={Jailbreak in Pieces: Compositional Adversarial Attacks on Multi-Modal Language Models},
  author={Shayegani, Erfan and Dong, Yue and Abu-Ghazaleh, Nael},
  journal={arXiv preprint arXiv:2307.14539},
  year={2023}
}

@article{zou2024poisonedrag,
  title={{PoisonedRAG}: Knowledge Poisoning Attacks to Retrieval-Augmented Generation of Large Language Models},
  author={Zou, Wei and Geng, Runpeng and Wang, Binghui and Jia, Jinyuan},
  journal={arXiv preprint arXiv:2402.07867},
  year={2024}
}

@article{chaudhari2024phantom,
  title={Phantom: General Trigger Attacks on Retrieval Augmented Language Generation},
  author={Chaudhari, Harsh and Abdelfattah, Giorgio and Perez, Ethan and Mitra, Subhabrata},
  journal={arXiv preprint arXiv:2405.20485},
  year={2024}
}

@article{bratus2017langsec,
  title={{LangSec}: Language-Theoretic Security},
  author={Bratus, Sergey and Locasto, Michael E and Patterson, Meredith L and Sassaman, Len and Shubina, Anna},
  journal={IEEE Security \& Privacy},
  volume={15},
  number={4},
  pages={36--42},
  year={2017},
  publisher={IEEE}
}

@inproceedings{sassaman2013security,
  title={Security Applications of Formal Language Theory},
  author={Sassaman, Len and Patterson, Meredith L and Bratus, Sergey and Locasto, Michael E},
  booktitle={IEEE Systems Journal},
  volume={7},
  number={3},
  pages={489--500},
  year={2013}
}

@inproceedings{momot2016langsec,
  title={The Seven Turrets of Babel: A Taxonomy of {LangSec} Errors and How to Expunge Them},
  author={Momot, Falcon and Bratus, Sergey and Hallberg, Sven M and Patterson, Meredith L},
  booktitle={IEEE Cybersecurity Development (SecDev)},
  pages={45--52},
  year={2016}
}

@article{saltzer1975protection,
  title={The Protection of Information in Computer Systems},
  author={Saltzer, Jerome H and Schroeder, Michael D},
  journal={Proceedings of the IEEE},
  volume={63},
  number={9},
  pages={1278--1308},
  year={1975},
  publisher={IEEE}
}

@book{anderson2020security,
  title={Security Engineering: A Guide to Building Dependable Distributed Systems},
  author={Anderson, Ross J},
  edition={3},
  year={2020},
  publisher={John Wiley \& Sons}
}

@article{denning1976lattice,
  title={A Lattice Model of Secure Information Flow},
  author={Denning, Dorothy E},
  journal={Communications of the ACM},
  volume={19},
  number={5},
  pages={236--243},
  year={1976},
  publisher={ACM}
}

@techreport{bell1973secure,
  title={Secure Computer Systems: Mathematical Foundations},
  author={Bell, David Elliott and LaPadula, Leonard J},
  institution={MITRE Corporation},
  number={MTR-2547},
  year={1973}
}

@inproceedings{provos2003preventing,
  title={Preventing Privilege Escalation},
  author={Provos, Niels and Friedl, Markus and Honeyman, Peter},
  booktitle={12th USENIX Security Symposium},
  pages={231--242},
  year={2003}
}

@article{alephone1996smashing,
  title={Smashing the Stack for Fun and Profit},
  author={{Aleph One}},
  journal={Phrack Magazine},
  volume={7},
  number={49},
  year={1996}
}

@inproceedings{cowan1998stackguard,
  title={{StackGuard}: Automatic Adaptive Detection and Prevention of Buffer-Overflow Attacks},
  author={Cowan, Crispin and Pu, Calton and Maier, Dave and Walpole, Jonathan and Bakke, Peat and Beattie, Steve and Grier, Aaron and Wagle, Perry and Zhang, Qian and Hinton, Heather},
  booktitle={7th USENIX Security Symposium},
  pages={63--78},
  year={1998}
}

@article{halfond2006classification,
  title={A Classification of {SQL}-Injection Attacks and Countermeasures},
  author={Halfond, William G and Viegas, Jeremy and Orso, Alessandro},
  journal={IEEE International Symposium on Secure Software Engineering},
  volume={1},
  pages={13--15},
  year={2006}
}

@inproceedings{vogt2007cross,
  title={Cross-Site Scripting Prevention with Dynamic Data Tainting and Static Analysis},
  author={Vogt, Philipp and Nentwich, Florian and Jovanovic, Nenad and Kirda, Engin and Kruegel, Christopher and Vigna, Giovanni},
  booktitle={Network and Distributed System Security Symposium (NDSS)},
  year={2007}
}

@inproceedings{vaswani2017attention,
  title={Attention Is All You Need},
  author={Vaswani, Ashish and Shazeer, Noam and Parmar, Niki and Uszkoreit, Jakob and Jones, Llion and Gomez, Aidan N and Kaiser, {\L}ukasz and Polosukhin, Illia},
  booktitle={Advances in Neural Information Processing Systems},
  volume={30},
  year={2017}
}

@article{brown2020language,
  title={Language Models Are Few-Shot Learners},
  author={Brown, Tom and Mann, Benjamin and Ryder, Nick and Subbiah, Melanie and Kaplan, Jared D and Dhariwal, Prafulla and Neelakantan, Arvind and Shyam, Pranav and Sastry, Girish and Askell, Amanda and others},
  journal={Advances in Neural Information Processing Systems},
  volume={33},
  pages={1877--1901},
  year={2020}
}

@article{ouyang2022training,
  title={Training Language Models to Follow Instructions with Human Feedback},
  author={Ouyang, Long and Wu, Jeffrey and Jiang, Xu and Almeida, Diogo and Wainwright, Carroll and Mishkin, Pamela and Zhang, Chong and Agarwal, Sandhini and Slama, Katarina and Ray, Alex and others},
  journal={Advances in Neural Information Processing Systems},
  volume={35},
  pages={27730--27744},
  year={2022}
}

@article{bai2022training,
  title={Training a Helpful and Harmless Assistant with Reinforcement Learning from Human Feedback},
  author={Bai, Yuntao and Jones, Andy and Ndousse, Kamal and Askell, Amanda and Chen, Anna and DasSarma, Nova and Drain, Dawn and Fort, Stanislav and Ganguli, Deep and Henighan, Tom and others},
  journal={arXiv preprint arXiv:2204.05862},
  year={2022}
}

@article{casper2023open,
  title={Open Problems and Fundamental Limitations of Reinforcement Learning from Human Feedback},
  author={Casper, Stephen and Davies, Xander and Shi, Claudia and Gilbert, Thomas Krendl and Scheurer, J{\'e}r{\'e}my and Rando, Javier and Freedman, Rachel and Korbak, Tomasz and Lindner, David and Freire, Pedro and others},
  journal={arXiv preprint arXiv:2307.15217},
  year={2023}
}

@article{wolf2023fundamental,
  title={Fundamental Limitations of Alignment in Large Language Models},
  author={Wolf, Yotam and Wies, Noam and Levine, Yoav and Shashua, Amnon},
  journal={arXiv preprint arXiv:2304.11082},
  year={2023}
}

@article{schick2023toolformer,
  title={Toolformer: Language Models Can Teach Themselves to Use Tools},
  author={Schick, Timo and Dwivedi-Yu, Jane and Dess{\`\i}, Roberto and Raileanu, Roberta and Lomeli, Maria and Zettlemoyer, Luke and Cancedda, Nicola and Scialom, Thomas},
  journal={arXiv preprint arXiv:2302.04761},
  year={2023}
}

@article{yao2023react,
  title={{ReAct}: Synergizing Reasoning and Acting in Language Models},
  author={Yao, Shunyu and Zhao, Jeffrey and Yu, Dian and Du, Nan and Shafran, Izhak and Narasimhan, Karthik and Cao, Yuan},
  journal={arXiv preprint arXiv:2210.03629},
  year={2023}
}

@article{xi2023rise,
  title={The Rise and Potential of Large Language Model Based Agents: A Survey},
  author={Xi, Zhiheng and Chen, Wenxiang and Guo, Xin and He, Wei and Ding, Yiwen and Hong, Boyang and Zhang, Ming and Wang, Junzhe and Jin, Senjie and Zhou, Enyu and others},
  journal={arXiv preprint arXiv:2309.07864},
  year={2023}
}

@article{wang2024survey,
  title={A Survey on Large Language Model Based Autonomous Agents},
  author={Wang, Lei and Ma, Chen and Feng, Xueyang and Zhang, Zeyu and Yang, Hao and Zhang, Jingsen and Chen, Zhiyuan and Tang, Jiakai and Chen, Xu and Lin, Yankai and others},
  journal={Frontiers of Computer Science},
  volume={18},
  number={6},
  pages={186345},
  year={2024}
}

@article{durante2024agent,
  title={Agent {AI}: Surveying the Horizons of Multimodal Interaction},
  author={Durante, Zane and Sarber, Bidipta and Gong, Ran and Tavassolipour, Rohan and Nishi, Kevin and Schettini, Riki and Navarro, Raul and Ceylan, Duygu and Chen, Xinpeng and Qu, Yujin and others},
  journal={arXiv preprint arXiv:2401.03568},
  year={2024}
}

@article{ruan2023identifying,
  title={Identifying the Risks of {LM} Agents with an {LM}-Emulated Sandbox},
  author={Ruan, Yangjun and Dong, Honghua and Wang, Andrew and Pitis, Silviu and Zhou, Yongchao and Ba, Jimmy and Dubois, Yann and Maddison, Chris J and Hashimoto, Tatsunori},
  journal={arXiv preprint arXiv:2309.15817},
  year={2023}
}

@article{weidinger2022taxonomy,
  title={Taxonomy of Risks Posed by Language Models},
  author={Weidinger, Laura and Mellor, Jonathan and Rauh, Maribeth and Griffin, Conor and Uesato, Jonathan and Huang, Po-Sen and Cheng, Myra and Glaese, Mia and Balle, Borja and Kasirzadeh, Atoosa and others},
  booktitle={ACM Conference on Fairness, Accountability, and Transparency},
  pages={214--229},
  year={2022}
}

@article{shavit2023practices,
  title={Practices for Governing Agentic {AI} Systems},
  author={Shavit, Yonatan and Agarwal, Sandhini and Brundage, Miles and Adler, Steven and O'Keefe, Cullen and Campbell, Rosie and Lee, Theodore and Mishkin, Pamela and Eloundou, Tyna and Hickey, Charlie and others},
  journal={OpenAI Research},
  year={2023}
}

@article{kinniment2024evaluating,
  title={Evaluating Language-Model Agents on Realistic Autonomous Tasks},
  author={Kinniment, Megan and Sato, Lucas Jun Koba and Du, Haoxing and Goodrich, Brian and Hasin, Max and Chan, Lawrence and Miles, Luke Harold and Lin, Tao and Wijk, Hjalmar and Burget, Joel and others},
  journal={arXiv preprint arXiv:2312.11671},
  year={2024}
}

@article{ioannidis2005most,
  title={Why Most Published Research Findings Are False},
  author={Ioannidis, John PA},
  journal={PLoS Medicine},
  volume={2},
  number={8},
  pages={e124},
  year={2005}
}

@article{baker2016reproducibility,
  title={1,500 Scientists Lift the Lid on Reproducibility},
  author={Baker, Monya},
  journal={Nature},
  volume={533},
  number={7604},
  pages={452--454},
  year={2016}
}

@article{wang2023scientific,
  title={Scientific Discovery in the Age of Artificial Intelligence},
  author={Wang, Hanchen and Fu, Tianfan and Du, Yuanqi and Gao, Wenhao and Huang, Kexin and Liu, Ziming and Chandak, Payal and Liu, Shengchao and Van Katwyk, Peter and Deac, Andreea and others},
  journal={Nature},
  volume={620},
  number={7972},
  pages={47--60},
  year={2023}
}

@article{birhane2023science,
  title={Science in the Age of Large Language Models},
  author={Birhane, Abeba and Kasirzadeh, Atoosa and Leslie, David and Wachter, Sandra},
  journal={Nature Reviews Physics},
  volume={5},
  number={5},
  pages={277--280},
  year={2023}
}

@article{moreau2013prov,
  title={{PROV-DM}: The {PROV} Data Model},
  author={Moreau, Luc and Missier, Paolo and others},
  journal={W3C Recommendation},
  year={2013}
}

@article{herschel2017survey,
  title={A Survey on Provenance: What For? What Form? What From?},
  author={Herschel, Melanie and Diestelk{\"a}mper, Ralf and Ben Lahmar, Houssem},
  journal={The VLDB Journal},
  volume={26},
  number={6},
  pages={881--906},
  year={2017}
}

@article{ganguli2022red,
  title={Red Teaming Language Models to Reduce Harms: Methods, Scaling Behaviors, and Lessons Learned},
  author={Ganguli, Deep and Lovitt, Liane and Kernion, Jackson and Askell, Amanda and Bai, Yuntao and Kadavath, Saurav and Mann, Ben and Perez, Ethan and Schiefer, Nicholas and Ndousse, Kamal and others},
  journal={arXiv preprint arXiv:2209.07858},
  year={2022}
}

@article{perez2022red,
  title={Red Teaming Language Models with Language Models},
  author={Perez, Ethan and Huang, Saffron and Song, Francis and Cai, Trevor and Ring, Roman and Aslanides, John and Glaese, Amelia and McAleese, Nat and Irving, Geoffrey},
  journal={arXiv preprint arXiv:2202.03286},
  year={2022}
}

@article{mazeika2024harmbench,
  title={{HarmBench}: A Standardized Evaluation Framework for Automated Red Teaming and Robust Refusal},
  author={Mazeika, Mantas and Phan, Long and Yin, Xuwang and Zou, Andy and Wang, Zifan and Mu, Norman and Sakhaee, Elham and Li, Nathaniel and Basart, Steven and Li, Bo and others},
  journal={arXiv preprint arXiv:2402.04249},
  year={2024}
}

@techreport{anderson1972computer,
  title={Computer Security Technology Planning Study},
  author={Anderson, James P},
  institution={Air Force Electronic Systems Division},
  number={ESD-TR-73-51},
  year={1972}
}

@inproceedings{jaeger2004consistency,
  title={Consistency Analysis of Authorization Hook Placement in the {Linux} Security Modules Framework},
  author={Jaeger, Trent and Edwards, Antony and Zhang, Xiaolan},
  booktitle={ACM Transactions on Information and System Security},
  volume={7},
  number={2},
  pages={175--205},
  year={2004}
}

@article{sandhu1996role,
  title={Role-Based Access Control Models},
  author={Sandhu, Ravi S and Coyne, Edward J and Feinstein, Hal L and Youman, Charles E},
  journal={Computer},
  volume={29},
  number={2},
  pages={38--47},
  year={1996}
}

@book{nipkow2002isabelle,
  title={Isabelle/{HOL}: A Proof Assistant for Higher-Order Logic},
  author={Nipkow, Tobias and Paulson, Lawrence C and Wenzel, Markus},
  year={2002},
  publisher={Springer}
}

@inproceedings{klein2009sel4,
  title={{seL4}: Formal Verification of an {OS} Kernel},
  author={Klein, Gerwin and Elphinstone, Kevin and Heiser, Gernot and Andronick, June and Cock, David and Derrin, Philip and Elkaduwe, Dhammika and Engelhardt, Kai and Kolanski, Rafal and Norrish, Michael and others},
  booktitle={ACM SIGOPS 22nd Symposium on Operating Systems Principles},
  pages={207--220},
  year={2009}
}

@inproceedings{cedar2024,
  title={Cedar: A New Language for Expressive, Fast, Safe, and Analyzable Authorization},
  author={Cutler, John and Hance, Travis and Headley, William and Ioannidis, Stratos and Parno, Bryan and Protzenko, Jonathan and Ramananandro, Tahina and Ringer, Aseem and Singh, Anmol and Wei, Aaron},
  booktitle={ACM SIGPLAN Conference on Object-Oriented Programming, Systems, Languages, and Applications (OOPSLA)},
  year={2024}
}

@article{damianou2001ponder,
  title={The {Ponder} Policy Specification Language},
  author={Damianou, Nicodemos and Dulay, Naranker and Lupu, Emil and Sloman, Morris},
  journal={International Workshop on Policies for Distributed Systems and Networks},
  pages={18--38},
  year={2001}
}

@article{boiko2023autonomous,
  title={Autonomous Chemical Research with Large Language Models},
  author={Boiko, Daniil A and MacKnight, Robert and Kline, Ben and Gomes, Gabe},
  journal={Nature},
  volume={624},
  number={7992},
  pages={570--578},
  year={2023}
}

@article{bran2023chemcrow,
  title={{ChemCrow}: Augmenting Large-Language Models with Chemistry Tools},
  author={Bran, Andres M and Cox, Sam and Schilter, Oliver and Baldassari, Carlo and White, Andrew D and Schwaller, Philippe},
  journal={arXiv preprint arXiv:2304.05376},
  year={2023}
}

\end{document}